\documentclass[aps,prl,twocolumn,superscriptaddress,showpacs,floatfix]{revtex4}
\usepackage{graphicx}
\usepackage{bm}
\begin{document}

\title{Conservation laws and thermodynamic efficiencies}

\author{Giuliano Benenti}
\affiliation{CNISM and Center for Nonlinear and Complex Systems,
Universit\`a degli Studi dell'Insubria, via Valleggio 11, 22100 Como, Italy}
\affiliation{Istituto Nazionale di Fisica Nucleare, Sezione di Milano,
via Celoria 16, 20133 Milano, Italy}

\author{Giulio Casati}
\affiliation{CNISM and Center for Nonlinear and Complex Systems,
Universit\`a degli Studi dell'Insubria, via Valleggio 11, 22100 Como, Italy}
\affiliation{Istituto Nazionale di Fisica Nucleare, Sezione di Milano,
via Celoria 16, 20133 Milano, Italy}

\author{Jiao Wang}
\affiliation{Department of Physics and Institute of Theoretical Physics
and Astrophysics, Xiamen University, Xiamen 361005, Fujian, China}

\date{\today}

\begin{abstract}
We show that generic systems with a single relevant conserved quantity
reach the Carnot efficiency in the thermodynamic limit. Such a general
result is illustrated by means of a diatomic chain of hard-point elastically
colliding particles where the total momentum is the only relevant conserved
quantity.
\end{abstract}

\maketitle

Conservation laws strongly affect transport properties. Conserved
quantities may lead to time correlations not decaying with time,
so that transport is not diffusive and is described, within the
linear response theory, by diverging transport coefficients.
This ideal conducting (ballistic) behavior can be firmly established
as a consequence of an inequality by Mazur~\cite{mazur,suzuki,prosen}
which, for a system of size $\Lambda$ characterized by $M$
conserved quantities $Q_n$, $n=1,\cdots,M$, bounds the time-averaged
current-current correlation functions as
\begin{equation}
\lim_{t\to\infty}\frac{1}{t}
\int_0^{t} dt' \langle J(t') J(0) \rangle_T \ge
\sum_{n=1}^{M} \frac{\langle J Q_n \rangle_T^2}{\langle Q_n^2\rangle_T},
\label{eq:mazur}
\end{equation}
where $\langle \cdots \rangle_T$ denotes the thermodynamic average
at temperature $T$. The constants of motion, $Q_n$, are orthogonal
to each other, i.e., $\langle Q_n Q_m \rangle_T = \langle Q_n^2 \rangle_T
\delta_{n,m}$, and \emph{relevant}, that is,
$\langle J Q_n \rangle_T\ne 0$ for all $n$.
A non-zero right-hand side in Eq.~(\ref{eq:mazur}) at the thermodynamic
limit implies a finite Drude weight for the current $J$, which in turn
indicates ballistic transport~\cite{zotos,zotosreview}. The impact of
motion constants on the electric and thermal conductivities has been
widely investigated~\cite{zotos,zotosreview,garst,heidrich-meisner}.
In particular, anomalous heat transport has been discussed for momentum
conserving interacting systems in low dimensions~\cite{lepri,dhar}.
However, to the best of our knowledge, conservation laws have never
been discussed for coupled flows, in particular in relation
to the problem of optimizing thermodynamic efficiencies.

The search of a new technology capable of reducing the environmental impact
of electrical power generation and refrigeration has aroused great interest
in thermoelectricity, namely the possibility to build a type of solid-state
heat engine capable of converting heat into electricity, or alternatively
electricity into cooling~\cite{majumdar,dresselhaus,snyder,dubi,shakuori}.
The main difficulty is connected to the low efficiency of such heat engine.
We recall that the maximum thermoelectric efficiency as well as the efficiency
at maximum power~\cite{vandenbroeck,esposito2009,schulman,esposito2010,seifert}
are determined, within the linear response regime and for systems with
time-reversal symmetry~\cite{footnote}, by the so called figure of merit $ZT$,
which is a dimensionless quantity, a combination of the three main
transport properties of a material: the thermal conductivity $\kappa$,
the electrical conductivity $\sigma$ and the thermopower
(Seebeck coefficient) $S$,
as well as of the absolute temperature $T$:
\begin{equation}
ZT=\frac{\sigma S^2}{\kappa}\, T.
\end{equation}
The maximum efficiency is given by
\begin{equation}
\eta_{\rm max}=
\eta_C\,
\frac{\sqrt{ZT+1}-1}{\sqrt{ZT+1}+1},
\label{etamaxB0}
\end{equation}
where $\eta_C$ is the Carnot efficiency,
while the efficiency
$\eta(W_{\rm max})$ at maximum output power
$W_{\rm max}$ reads~\cite{vandenbroeck}
\begin{equation}
\eta(W_{\rm max})=\frac{\eta_{C}}{2}\frac{ZT}{ZT+2}.
\label{etawmaxB0}
\end{equation}
The only restriction imposed by thermodynamics is $ZT\ge 0$, so that
both efficiencies are monotonous growing functions of the figure of merit
and $\eta_{\rm max}\to \eta_C$, $\eta(W_{\rm max})\to
\frac{\eta_C}{2}$ when $ZT\to\infty$.
It has been suggested that the value $ZT=3$ is the target to be achieved
in order to make thermoelectric engines economically competitive. In spite
of recent progress in material science, present technology is limited to low
$ZT$ materials and no clear path has been identified in order to increase
efficiency.

A promising new approach, based on the theory of dynamical systems, has
been recently introduced~\cite{prlCMP,reviewZT}. The hope is that the
analysis of idealized models may lead to some insight on the microscopic
mechanisms which lead to high figure of merit in more realistic materials.
While for non-interacting systems, even in the classical framework, the
energy filtering mechanism~\cite{mahansofo,linke1,linke2} has been shown
to allow to reach Carnot efficiency, very little is known for interacting
particles. Recent numerical and empirical evidence has shown that for a
one-dimensional diatomic disordered chain of hard-point elastically
colliding particles, the figure of merit $ZT$ diverges as the number of
particles increases~\cite{Casati09,engine09}. Since it has been verified
that the energy filtering mechanism does not work here~\cite{Keiji10},
it follows that the divergence of $ZT$ in the thermodynamic limit rests
on a different, unknown property.

In the present paper we analyze and solve this problem. Indeed, we show
that for systems having a single relevant constant of motion, the electric
conductivity is ballistic, i.e., $\sigma\propto \Lambda$, the heat
conductivity is subballistic, $\kappa \propto \Lambda^\alpha$ with $\alpha<1$,
and the thermopower is size independent, $S\propto \Lambda^0$, so that
the figure of merit $ZT\propto \Lambda^{1-\alpha} \to \infty$ in the
thermodynamic limit $\Lambda\to\infty$. Our findings are illustrated by
the above mentioned prototype model of interacting one-dimensional system:
a diatomic chain of hard-point elastically colliding particles, where
the total momentum is the only relevant constant of motion.

We start from the equations connecting fluxes and thermodynamic
forces within linear irreversible thermodynamics~\cite{callen,degrootmazur}:
\begin{equation}
\left\{
\begin{array}{l}
{\displaystyle
J_\rho=L_{\rho\rho} X_1 + L_{\rho u} X_2,
}
\\
\\
{\displaystyle
J_u=L_{u\rho} X_1 + L_{uu} X_2,
}
\end{array}
\right.
\end{equation}
where $J_\rho$ and $J_u$ are the particle and energy currents, and the
thermodynamic forces $X_1=-\nabla(\beta\mu)$, $X_2=\nabla \beta$, with
$\mu$ chemical potential and $\beta=1/T$ inverse temperature. (We set
the Boltzmann constant $k_B=1$.) The Onsager coefficients $L_{ij}$
($i,j=\rho,u$) are related to the familiar transport coefficients as
follows:
\begin{equation}
\sigma=\frac{L_{\rho\rho}}{T},
\quad\kappa=\frac{1}{T^2}\frac{\det {\bm L}}{L_{\rho\rho}},
\quad S=\frac{1}{T}\left(\frac{L_{\rho u}}{L_{\rho\rho}}-\mu\right),
\end{equation}
where ${\bm L}$ denotes the Onsager matrix with matrix elements
$L_{ij}$ and we have set the electric charge 
of each particle $e=1$. Thermodynamics
imposes $\det {\bm L}\ge 0$, $L_{\rho\rho}\ge 0$, $L_{uu}\ge 0$,
and $L_{u\rho}=L_{\rho u}$. The figure of merit reads
\begin{equation}
ZT=\frac{(L_{u\rho}-\mu L_{\rho\rho})^2}{\det {\bm L}}.
\end{equation}
It diverges (thus leading to maximum efficiency) iff the Onsager matrix
${\bm L}$ is ill-conditioned, that is, in the so-called strong-coupling
condition, for which the energy and particle currents are proportional,
$J_u\propto J_\rho$, the proportionality factor being independent of the
values of the applied thermodynamic forces.

The Green-Kubo formula expresses the Onsager coefficients in terms
of correlation functions of the corresponding current operators,
calculated at thermodynamic equilibrium~\cite{kubo,mahan}:
\begin{equation}
L_{ij} = \lim_{\omega\to 0} {\rm Re} L_{ij} (\omega),
\end{equation}
where
\begin{equation}
\begin{array}{c}
L_{ij}(\omega)\equiv \lim_{\epsilon\to 0}
\int_0^\infty dt e^{-i(\omega-i\epsilon)t}
\\
\\
\times
\lim_{\Lambda \to\infty}
\frac{1}{\Lambda}
\int_0^\beta d\tau\langle {J}_i {J}_j (t+i\tau)\rangle_T.
\end{array}
\label{eq:kubo}
\end{equation}
The real part of $L_{ij}(\omega)$ can be decomposed into a $\delta$-function
at zero frequency defining the \emph{generalized Drude weight}
${\cal D}_{ij}$ (for $i=j=\rho$ this is the conventional Drude weight) and
a regular part $L_{ij}^{\rm reg}(\omega)$:
\begin{equation}
{\rm Re} L_{ij}(\omega)=
2\pi {\cal D}_{ij}\delta(\omega)+L_{ij}^{\rm reg}(\omega).
\label{eq:reLij}
\end{equation}
Non-zero Drude weights, ${\cal D}_{ij}\ne 0$  for $i,j=\rho,u$ are a
signature of ballistic transport, namely $L_{ij}\propto \Lambda$ at
the thermodynamic limit, and
therefore the thermopower
$S=L_{\rho u}/(TL_{\rho\rho})-\mu/T\propto \Lambda^0$.

We now discuss the influence of conserved quantities on
the figure of merit $ZT$. We make use of Suzuki's
formula~\cite{suzuki} for the currents $J_\rho$ and $J_u$,
which generalizes Mazur's inequality (\ref{eq:mazur}) by stating that,
for a system of finite size $\Lambda$,
\begin{equation}
\begin{array}{c}
{\displaystyle
C_{ij}(\Lambda)\equiv
\lim_{t\to\infty}\frac{1}{t}
\int_0^{t} dt' \langle J_i(t') J_j(0) \rangle_T
}
\\
\\
{\displaystyle
=
\sum_{n=1}^M
\frac{\langle J_i Q_n \rangle_T
\langle J_j Q_n \rangle_T}{\langle Q_n^2\rangle_T},
}
\end{array}
\label{eq:suzuki}
\end{equation}
where the summation is extended over all the $M$ orthogonal
constants of motion which are relevant for the considered flows,
that is, non-orthogonal to the currents $J_\rho$ and $J_u$,
i.e., $\langle J_\rho Q_n \rangle_T\ne 0$ and
$\langle J_u Q_n \rangle_T\ne 0$.
(Irrelevant constants of motion are not included in the
summation since they would give zero contribution.) The presence
of relevant conservation laws implies that the \emph{finite-size}
generalized Drude weights
\begin{equation}
D_{ij}(\Lambda)\equiv \frac{1}{2\Lambda}\,C_{ij}(\Lambda)
\end{equation}
are different from zero.
If at the thermodynamic limit the generalized Drude weight
\begin{equation}
{\cal D}_{ij}=\lim_{t\to\infty}\lim_{\Lambda\to \infty}
\frac{1}{2\Lambda t}
\int_0^{t} dt' \langle J_i(t') J_j(0) \rangle_T
\label{eq:drudeinfinite}
\end{equation}
is non-zero, then we can conclude that transport is 
ballistic~\cite{footnote4}.
Note that in Eq.~(\ref{eq:drudeinfinite}) the thermodynamic limit
$\Lambda\to\infty$ must be taken before the long-time limit $t\to\infty$.
The below developed theory only applies to the cases in which the two
limits commute, that is,
${\cal D}_{ij}=\lim_{\Lambda\to\infty} D_{ij}(\Lambda)$~\cite{footnote2}.

If there is a \emph{single} relevant constant of motion, $M=1$, due to
Suzuki's formula (\ref{eq:suzuki}) (and assuming that the two limits
$\Lambda\to\infty$ and $t\to 0$ commute), the ballistic contribution
to $\det {\bm L}$ vanishes, since it is proportional to
${\cal D}_{\rho\rho}{\cal D}_{uu}-{\cal D}_{\rho u}^2$, which is zero
from (\ref{eq:suzuki}). Hence, $\det {\bm L}$ grows due to the
contributions involving the regular part in Eq. (\ref{eq:reLij}), 
i.e., slower
than $\Lambda^2$, thus implying that the thermal conductivity
$\kappa\propto \det{\bm L}/L_{\rho\rho}$ grows subballistically.
That is, $\kappa\propto \Lambda^\alpha$, with $\alpha<1$.
Since $\sigma\propto L_{\rho\rho}$ is ballistic and
$S=L_{\rho u}/TL_{\rho\rho}-\mu/T\propto \Lambda^0$,
we can conclude that $ZT=\sigma S^2 T/\kappa
\propto \Lambda^{1-\alpha}\to\infty$ when $\Lambda\to\infty$.

The situation is drastically different
if $M>1$. In this case, due to the Schwartz inequality,
\begin{equation}
D_{\rho\rho}D_{uu}-D_{\rho u}^{2}=
||{\bm x}_\rho||^2 ||{\bm x}_u||^2-
\langle {\bm x}_\rho , {\bm x}_u \rangle \ge 0,
\end{equation}
where
\begin{equation}
{\bm x}_i= (x_{i1},...,x_{iM})=\frac{1}{2\Lambda}
\left(\frac{\langle J_i Q_1\rangle_T}{\sqrt{\langle Q_1^2 \rangle_T}},...,
\frac{\langle J_i Q_M\rangle_T}{\sqrt{\langle Q_M^2 \rangle_T}}
\right),
\end{equation}
and $\langle {\bm x}_\rho , {\bm x}_u \rangle = \sum_{k=1}^M
x_{\rho k} x_{uk}$.
The equality arises only in the exceptional case when the vectors
${\bm x}_{\rho}$ and ${\bm x}_u$ are parallel.
Hence, for $M>1$ we expect, in general, $\det {\bm L}\propto \Lambda^2$,
so that heat transport is ballistic
and $ZT\propto \Lambda^0$.

In order to illustrate the above general ideas, we consider a
one-dimensional, diatomic disordered chain of $N$ hard-point elastically
colliding particles with randomly distributed coordinates $z_i\in [0,\Lambda]$, velocities $v_i$, and masses $m_i\in \{\nu_1,\nu_2\}$ . The numerically
observed divergence of the figure of merit $ZT$ for
$\nu_1\ne \nu_2$~\cite{Casati09,engine09,Keiji10} can be understood now
in terms of the above developed theory. Indeed, in this system there
is a single relevant constant of motion $Q_1=P$, where
$P=\sum_{i=1}^N m_i v_i$ is the overall momentum~\cite{footnote3}. In this
case the particle current $J_\rho=\sum_{i=1}^{N} v_i$ and the energy current
$J_u=\sum_{i=1}^{N} \frac{1}{2}\, m_i v_i^3$. Note that the mass current,
$J_m\equiv\sum_{i=1}^N m_i v_i$, equals the total momentum $P$ and therefore
does not decay, while on the other hand $J_m\sim \bar{m} J_\rho$, where
$\bar{m}$ is the average mass per particle, hence we expect that
$\langle J_\rho(t)J_\rho(0)\rangle$ does not decay either, so that
$\sigma\sim \Lambda$.

It is easy to compute analytically the time-averaged correlation functions
$C_{ij}(\Lambda)$ (from the second line of Eq.~(\ref{eq:suzuki})) and then
the finite-size generalized Drude weights
\begin{equation}
\begin{array}{c}
{\displaystyle
D_{\rho\rho}(\Lambda)=\frac{C_{\rho\rho}(\Lambda)}{2\Lambda}=
\frac{T N^2}{2\Lambda(\nu_1 N_1 + \nu_2 N_2)},
}
\\
\\
{\displaystyle
D_{uu}(\Lambda)=\frac{C_{uu}(\Lambda)}{2\Lambda}=
\frac{9 T^3 N^2}{8\Lambda(\nu_1 N_1 + \nu_2 N_2)},
}
\\
\\
{\displaystyle
D_{\rho u}(\Lambda)=\frac{C_{\rho u}(\Lambda)}{2\Lambda}=
\frac{3 T^2 N^2}{4\Lambda(\nu_1 N_1 + \nu_2 N_2)}.
}
\end{array}
\label{eq:Drudehardpoint}
\end{equation}
Here $N=N_1+N_2$ with $N_1$ and $N_2$ the number of particles with
mass $\nu_1$ and $\nu_2$, respectively. Note that, as expected from
the above theory,
$D_{\rho\rho}(\Lambda)D_{uu}(\Lambda)-D_{\rho u}^2(\Lambda)=0$
for any system size $\Lambda$.

\begin{figure}
\vspace{0.cm}
\includegraphics[width=.99\columnwidth,clip]{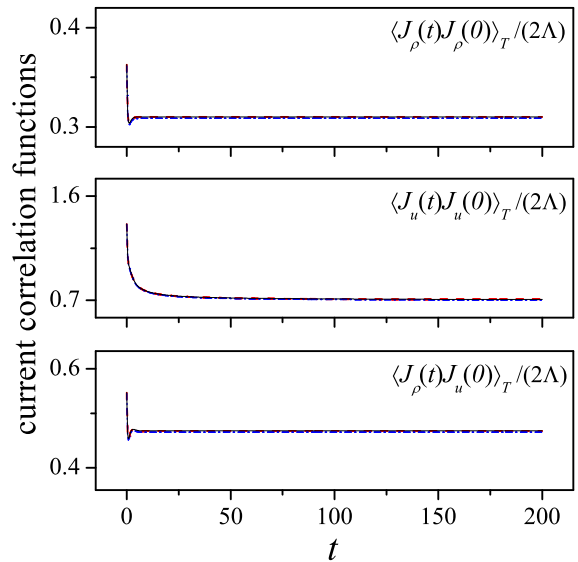}\vspace{-0.4cm}
\caption{(Color online) The current-current correlation functions for various system
sizes with $\Lambda=256$ (red dashed curve), $512$ (blue dash-dotted
curve), and $1024$ (black solid curve). The temperature $T=1$, particle
masses are $\nu_1=1$ and $\nu_2=\sqrt{5}$. It is seen that all the
current-current correlation functions approach a finite nonzero value
as the correlation time increases and that the characteristic time
scale to approach such value is independent of the system size.
Note that in all the figures the particle density is
fixed to be $N/\Lambda=1$ and $N_1=N_2=N/2$.}
\label{fig:correlations}
\end{figure}

\begin{figure}[!ht]
\includegraphics[width=.99\columnwidth,clip]{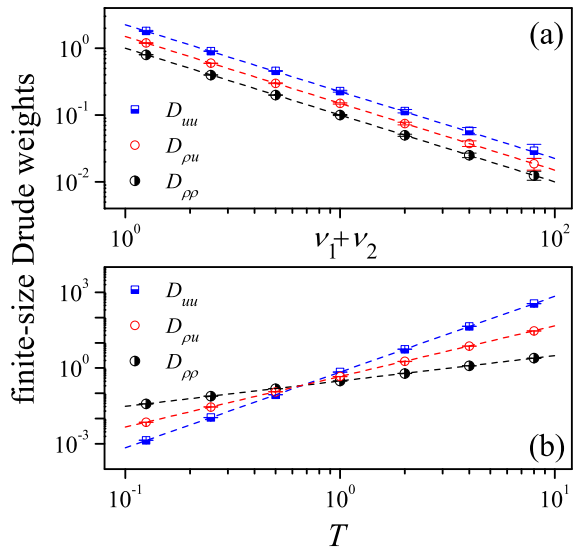}\vspace{-0.4cm}
\caption{(Color online) Comparison between the numerically determined finite-size generalized
Drude weights (symbols) and the analytical values for $D_{ij}$ (lines) given
in Eq.~(\ref{eq:Drudehardpoint}) for $\Lambda=256$. We set $T=1$ and
$\nu_1=1$ in (a) and $\nu_1=1$, $\nu_2=\sqrt{5}$ in (b).}
\label{fig:drude}
\end{figure}

To numerically confirm the above results, we compute the autocorrelation
functions of $J_\rho$ and $J_u$, and the cross correlation function between
them. In doing so we apply periodic boundary conditions and assign to
$N_1$ ($N_2$) particles of mass $\nu_1$ ($\nu_2$) random initial positions
and random initial velocities derived from the Maxwell-Boltzmann distribution corresponding to temperature $T$. We then evolve the system and compute
$c_{ij}(\Lambda,t) =
\frac{1}{t}\int_0^t dt' \langle J_i (t') J_j (0)\rangle_T$ (for $i,j=\rho, u$)
up to a time $t$ sufficiently long to
obtain stable averages, thus estimating
$C_{ij}(\Lambda)=\lim_{t\to\infty} c_{ij}(\Lambda,t)$.
We show in Fig.~\ref{fig:correlations}
that the current-current correlation functions
$\langle J_i (t) J_j (0)\rangle_T$
do not decay to zero as the correlation time is increased,
implying that $c_{ij}(\Lambda,t)$
do not decay either and thus indicating ballistic transport.
We finally estimate the finite-size generalized Drude weights
from the time-averaged correlation functions as
$c_{ij}(\Lambda,t)/(2\Lambda)$,
with a sufficiently long time $t$ to approximate
the asymptotic value $D_{ij}(\Lambda)=C_{ij}(\Lambda)/(2\Lambda)$.
As shown in Fig.~\ref{fig:drude}, the numerically
determined values are in very good agreement
with the theoretical values $D_{ij}$ given
by Eq.~(\ref{eq:Drudehardpoint}).

Fig.~\ref{fig:correlations} provides clear evidence that
the convergence of the correlation functions $c_{ij}(\Lambda,t)$
to their asymptotic values $C_{ij}(\Lambda)$ takes place on a time
scale independent of the system size $\Lambda$.
Therefore, Fig.~\ref{fig:correlations} provides a strong indication
that for the model under investigation the two
limits $\Lambda\to\infty$ and $t\to \infty$ do commute, so that
we can compute  the generalized Drude weights at the thermodynamic
limit as ${\cal D}_{ij} = \lim_{\Lambda\to \infty} D_{ij}(\Lambda)$.
Note that in taking the thermodynamic limit, we keep constant
the particle density $N/\Lambda$ and the ratio $N_1/N_2$.

\begin{figure}[ht]
\vspace{0.cm}
\includegraphics[width=.99\columnwidth,clip]{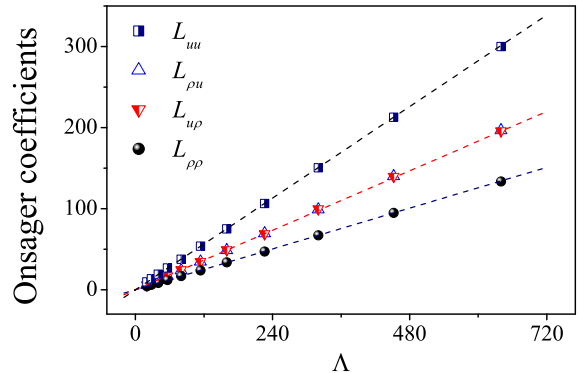}\vspace{-0.4cm}
\caption{(Color online) Dependence of the Onsager coefficients $L_{ij}$ on the
system size $\Lambda$. The straight lines are the best linear
fitting, $L_{ij}\propto \Lambda$. In these nonequilibrium simulations
we set the temperature $T=1$, the chemical potential $\mu=0$, and
the masses $\nu_1=1$ and $\nu_2=\sqrt{19}$.}
\label{fig:onsager}
\end{figure}

\begin{figure}[ht]
\vspace{0.cm}
\includegraphics[width=.99\columnwidth,clip]{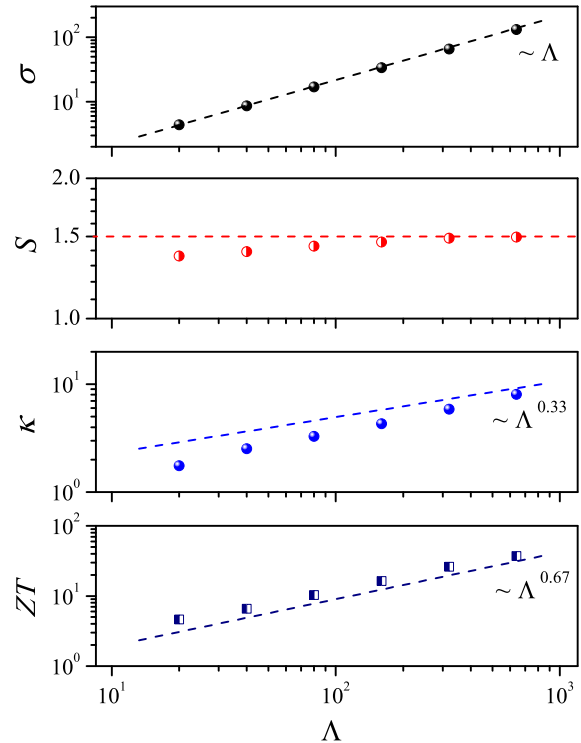}\vspace{-0.4cm}
\caption{(Color online) Dependence of the transport coefficients $\sigma$, $S$, $\kappa$
and $ZT$ on the system size $\Lambda$, for the same parameter values as in Fig.~\ref{fig:onsager}. Dashed lines are drawn for reference.}
\label{fig:ZT}
\end{figure}

Finally, we perform a nonequilibrium calculation of the various
transport coefficients. (For technical details of numerical simulations
see Ref.~\cite{engine09}.) According to our theory, we expect that all
the Onsager coefficients grow linearly with the system size in the
thermodynamic limit, since the generalized Drude weights ${\cal D}_{ij}$
are all different from zero. This expectation is confirmed by the data
shown in Fig.~\ref{fig:onsager}. Finally, in Fig.~\ref{fig:ZT} we show
the transport coefficients $\sigma$, $S$, $\kappa$, and the thermoelectric
figure of merit $ZT$ as a function of the system size. In agreement
with our theory, we observe that $\sigma\propto \Lambda$, while $S$
saturates to the value predicted from theory for ballistic transport,
$S=\frac{1}{T}\left(\frac{{\cal D}_{\rho u}}{{\cal D}_{\rho\rho}}
-\mu\right)=\frac{3}{2}$, $\kappa \propto \Lambda^\alpha$ with $\alpha
\approx 1/3$ \cite{Narayan, lepri1}, and the growth of the figure of merit 
in good agreement with the dependence $ZT\propto \Lambda^{1-\alpha}$.

Note that the above conclusions for the thermal conductivity $\kappa$
and the figure of merit $ZT$ do not hold in the integrable case
$\nu_1=\nu_2$, where $M>1$ since all moments of the momentum distribution
are conserved quantities. In this case $\kappa\propto \Lambda$ and it
is easy to analytically compute $ZT=1$~\cite{Casati09}.

To summarize, we have shown that for systems with a \emph{single
relevant} constant of motion, the thermoelectric figure of merit
diverges as the system size increases, so that the Carnot efficiency
is achieved at the thermodynamic limit. Such a result has
been illustrated in the case of a chain of hard-point elastically
colliding particles, with a remarkable agreement between analytical results,
equilibrium and out-of-equilibrium numerical simulations. We would like
to point out that, while our illustrative model is one-dimensional, there are
no dimensionality restrictions in our theory, so that it should apply
also to two- and three-dimensional systems in which total momentum is the
only relevant constant of motion. Therefore, our paper unveils a rather
generic mechanism for increasing thermoelectric efficiency in
interacting systems.

Useful discussions with Stefano Lepri are gratefully acknowledged.
G.B. and G.C. acknowledge the support by
MIUR-PRIN 2008 and by Regione Lombardia,
J.W. acknowledges the support by the NNSF (Grant No. 11275159)
and SRFDP (Grant No. 20100121110021) of China.


\begin{thebibliography}{99}

\bibitem{mazur}
P. Mazur, Physica (Amsterdam) {\bf 43}, 533 (1969).

\bibitem{suzuki}
M. Suzuki, Physica (Amsterdam) {\bf 51}, 277 (1971).

\bibitem{prosen}
E. Ilievski and T. Prosen, preprint arXiv:1111.3830 [math-ph],
Commun. Math. Phys. (in press).

\bibitem{zotos}
X. Zotos, F. Naef, and P. Prelov\v sek,
Phys. Rev. B {\bf 55}, 11029 (1997).

\bibitem{zotosreview}
X. Zotos and P. Prelov\v sek, in
D. Baeriswyl and L. Degiorgi (Eds.),
{\it Strong Interactions in Low Dimensions.}
(Kluwer Academic Publishers, Dordrecht. 2004).

\bibitem{garst}
M. Garst and A. Rosch,
Europhys. Lett. {\bf 55}, 66 (2001).

\bibitem{heidrich-meisner}
F. Heidrich-Meisner, A. Honecker, and W. Brenig,
Phys. Rev. B {\bf 71}, 184415 (2005).

\bibitem{lepri}
S. Lepri, R. Livi, and A. Politi,
Phys. Rep. {\bf 377}, 1 (2003).

\bibitem{dhar}
A. Dhar,
Adv. Phys. {\bf 57}, 457 (2008).


\bibitem{majumdar} A. Majumdar, Science {\bf 303}, 777 (2004).

\bibitem{dresselhaus}
M. S. Dresselhaus, G. Chen, M. Y. Tang, R. G. Yang, H. Lee,
D. Z. Wang, Z. F. Ren, J. -P. Fleurial, and P. Gogna,
Adv. Mater. {\bf 19}, 1043 (2007).

\bibitem{snyder}
G. J. Snyder and E. S. Toberer,
Nature Mater. {\bf 7}, 105 (2008).

\bibitem{dubi}
Y. Dubi and M. Di Ventra,
Rev. Mod. Phys. {\bf 83}, 131 (2011).

\bibitem{shakuori}
A. Shakouri,
Annu. Rev. Mater. Res. {\bf 41}, 399 (2011).

\bibitem{vandenbroeck}
C. Van den Broeck,
Phys. Rev. Lett. {\bf 95}, 190602 (2005).

\bibitem{esposito2009}
M. Esposito, K. Lindenberg, and C. Van den Broeck,
Phys. Rev. Lett. {\bf 102}, 130602 (2009).

\bibitem{schulman}
B. Gaveau, M. Moreau, and L.S. Schulman,
Phys. Rev. Lett. {\bf 105}, 060601 (2010).

\bibitem{esposito2010}
M. Esposito, R. Kawai, K. Lindenberg, and C. Van den Broeck,
Phys. Rev. Lett. {\bf 105}, 150603 (2010).

\bibitem{seifert}
U. Seifert,
Phys. Rev. Lett. {\bf 106}, 020601 (2011).

\bibitem{footnote}
Thermodynamic bounds on efficiency for systems with broken time-reversal
symmetry are discussed in Ref.~\cite{benenti2011}.

\bibitem{benenti2011}
G. Benenti, K. Saito, and G. Casati,
Phys. Rev. Lett.{\bf 106}, 230602 (2011).

\bibitem{prlCMP}
G. Casati, C. Mej\'{\i}a-Monasterio, and T. Prosen,
Phys. Rev. Lett. {\bf 98}, 104302 (2007).

\bibitem{reviewZT}
G. Benenti and G. Casati,
Phil. Trans. R. Soc. A {\bf 369}, 466 (2011).


\bibitem{mahansofo}
G. D. Mahan and J. O. Sofo,
Proc. Natl. Acad. Sci. USA {\bf 93}, 7436 (1996).

\bibitem{linke1}
T. E. Humphrey, R. Newbury, R. P. Taylor, and H. Linke,
Phys. Rev. Lett. {\bf 89} 116801 (2002).

\bibitem{linke2}
T.E. Humphrey and H. Linke,
Phys. Rev. Lett. {\bf 94} 096601 (2005).

\bibitem{Casati09} G. Casati, L. Wang, and T. Prosen,
J. Stat. Mech., L03004 (2009).

\bibitem{engine09} J. Wang, G. Casati, T. Prosen, and C.-H. Lai,
Phys. Rev. E {\bf 80}, 031136 (2009).

\bibitem{Keiji10} K. Saito, G. Benenti, and G. Casati,
Chem. Phys. {\bf 375}, 508 (2010).

\bibitem{callen}
H. B. Callen, {\it Thermodynamics and an Introduction to Thermostatics}
(second edition) (John Wiley \& Sons, New York, 1985).

\bibitem{degrootmazur}
S. R. de Groot and P. Mazur, {\it Nonequilibrium Thermodynamics}
(North-Holland, Amsterdam, 1962).

\bibitem{kubo}
R. Kubo, M. Toda, and N. Hashitsume,
{\it Statistical Physics II:
Nonequilibrium Statistical Mechanics}
(Springer-Verlag, 1985).

\bibitem{mahan}
G. D. Mahan, {\it Many-Particle Physics}
(Plenum Press, New York, 1990).

\bibitem{footnote4}
See Ref.~\cite{prosen} for a detailed discussion and derivation 
of Eq.~(\ref{eq:drudeinfinite}).

\bibitem{footnote2}
See Ref.~\cite{prosen} for a proof of the commutation of the
two limits for a class of quantum spin chains.

\bibitem{footnote3}
Total energy $E$ and number of particles $N_1$ and $N_2$ are also
constants of motion. However, theys are not relevant since they
are even functions of the velocities $v_i$ and therefore the
thermodynamic averages $\langle EJ_i\rangle_T$,
$\langle N_1 J_i\rangle_T$, and $\langle  N_2 J_i\rangle_T$
($i=\rho,u$) vanish, being $J_\rho$ and $J_u$ odd functions
of velocities.

\bibitem{Narayan}
O. Narayan and S. Ramaswamy, Phys. Rev. Lett. {\bf 89}, 200601 (2002).

\bibitem{lepri1}
L. Delfini, S. Lepri, R. Livi, and A. Politi,
Phys. Rev. E {\bf 73}, 060201(R) (2006); J. Stat. Mech. P02007 (2007).


\end{thebibliography}
\end{document}